\documentstyle[prl,aps]{revtex}

\input epsf
\epsfverbosetrue

\begin{document}
\renewcommand{\textfraction}{0.02}

\twocolumn[\hsize\textwidth\columnwidth\hsize\csname
@twocolumnfalse\endcsname
\title{Electronic structure of single-crystalline Mg$_x$Al$_{1-x}$B$_2$}
\author{S. Schuppler,$^{1}$ E. Pellegrin,$^{1}$ N. N\"ucker,$^{1}$
T. Mizokawa,$^{2}$ M. Merz,$^{3}$ D. A. Arena,$^{4}$ J.
Dvorak,$^{5}$ Y. U. Idzerda,$^{5}$ D.-J. Huang,$^{6}$ C.-F.
Cheng,$^{6}$ K.-P. Bohnen,$^{1}$ R. Heid,$^{1}$ P.
Schweiss,$^{1}$ and Th. Wolf$^{1}$}

\address{
$^1$ Forschungszentrum Karlsruhe, Institut f\"ur Festk\"orperphysik,
D-76021 Karlsruhe, Germany\\
$^2$ Department of Complexity Science and Engineering, University
of Tokyo, Tokyo 113-0033, Japan\\
$^3$ Institut f\"ur Kristallographie, Rheinisch-Westf\"alische Technische
Hochschule Aachen, D-52056 Aachen, Germany\\
$^4$ Naval Research Laboratory, Code 6345, Washington DC 20365,
USA\\
$^5$ Department of Physics, Montana State University, Bozeman, MT 59717, USA\\
$^6$ Synchrotron Radiation Research Center, No.\ 1 R\&D Road VI,
Hsinchu 30077, Taiwan}
\date{Received May 10, 2002}
\maketitle

\begin{abstract}
Polarization-dependent x-ray absorption spectroscopy at the B 1$s$
edge of single-crystalline Mg$_x$Al$_{1-x}$B$_2$ reveals a
strongly anisotropic electronic structure near the Fermi energy.
Comparing spectra for superconducting compounds ($x$=0.9, 1.0)
with those for the non-superconductor $x$=0.0 gives direct
evidence on the importance of an in-plane spectral feature
crossing $E_F$ for the superconducting properties of the
diborides. Good agreement is found with the projected B 2$p$
density of states from LDA band structure calculations.
\end{abstract}

\pacs{PACS numbers: 74.25.Jb, 71.20.Lp, 78.70.Dm}]

MgB$_2$ is a compound known for decades which only recently, to
everyone's surprise, was shown to be a superconductor
\cite{akimitsu01} with $T_c\approx39$ K, one of the highest
transition temperatures observed for any non-cuprate material.
Compared to the high-$T_c$ cuprates its crystal structure is
simple: boron and magnesium form separate layers (``in-plane''
lattice constant $a\approx3.09$ {\AA}) which, for B, are
graphite-like with a hexagonal atomic arrangement. The B and Mg
sheets are stacked alternatingly along the $c$ axis
(``out-of-plane'' lattice constant $c\approx3.52$ {\AA})\@. This
layered structure, and especially the covalent $sp^2$ and 2D
character of the B-B $\sigma$ bonds \cite{an01}, already suggests
a strong anisotropy between in-plane and out-of-plane properties
\cite{hue}\@. It was established quickly \cite{budko01,bohnen01}
by experiment and theory that superconductivity in this compound
is fairly conventional, being $s$-wave, BCS-type, and
phonon-induced. Recently, however, also deviations were
suggested, like the presence of two energy gap scales
\cite{twogap}\@. Substitution of Mg by Al is, at high
temperatures, possible for all concentrations, preserving the
crystal structure while decreasing lattice constants (slightly in
the case of $a$, considerably in the case of $c$ \cite{pena02});
electron doping, combined with increased orbital overlap due to
unit volume contraction, is expected during this substitution, and
superconductivity is very quickly suppressed at only modest
substitution levels \cite{slusky01}.

The electronic structure near the Fermi level, $E_F$, has been
studied by photoemisssion, x-ray absorption, and optical
experiments, first on polycrystalline samples
\cite{takahashi01,callcott011,callcott012,gorshunov01}, and now
augmented by de Haas--van Alphen investigations \cite{dHvA}\@.
Single crystals which recently became available \cite{lee01} allow
studies in even greater depth, like determining occupied band
dispersions in angle-resolved photoemission \cite{shen01}\@. More
immediate access to the {\em orbital} character is provided by
polarization-dependent near-edge x-ray absorption fine structure
(NEXAFS) experiments; we present the first such study on
single-crystalline Mg$_x$Al$_{1-x}$B$_2$, which, performed at the
boron 1$s$ edge, yields the unoccupied electronic structure with
B 2$p$ character (as a function of Al substitution) and is able
to discriminate between $\sigma$ and $\pi$ states. Spectral
results are directly compared with the appropriately projected B
2$p$ density of electronic states (DOS) from LDA band structure
calculations, enabling us to sensitively test and possibly
cross-corroborate the theoretical and experimental electronic
structure near $E_F$.

Single crystals of AlB$_2$ and the solid solution
Mg$_{0.9}$Al$_{0.1}$B$_2$ were grown from Al or Al/Mg flux by the
slow-cooling method; the resulting crystal sizes were about
2$\times$2 mm$^2$ for $x$=0 and 0.4$\times$0.5 mm$^2$ for
$x$=0.9\@. Hexagonal single-crystalline MgB$_2$ platelets with a
diameter of up to 200 $\mu$m could be obtained by isothermally
annealing $^{11}$B powder \cite{B11} in a Mg flux enclosed in an
evacuated Mo cylinder. The MgB$_2$ crystals ($x$=1) show a sharp
superconducting transition at $T_c$=38.8 K, with a resistive
10\%-90\% width $\Delta$$T$$\approx$1.8 K\@. The substitution
level, $x$=0.9, of the partially Al substituted crystals was
determined by x-ray diffraction; the transition was already
suppressed to around 20 K and considerably broadened. To increase
useful surface area for $x$=0.9 and 1.0, several crystals were
mounted right next to each other on the sample holder
\cite{orient}\@.

Near-edge x-ray absorption fine structure (NEXAFS) measurements
with linearly polarized light were performed at two facilities:
the National Synchrotron Light Source (NSLS), Brookhaven, USA,
where the Montana State University/NSLS beamline U4B was utilized
for measuring the $x$=0 sample; and at the Synchrotron Radiation
Research Center (SRRC), Hsinchu, Taiwan, where the undulator
beamline U5 was used for measuring the $x$=0.9 and $x$=1.0
samples. Resolution was set to 180 and 100 meV at the NSLS and
the SRRC, resp.; all data were taken in the normal state.
Cleaving the samples was obviated by employing bulk-sensitive
fluorescence-yield (FY)
\twocolumn[\hsize\textwidth\columnwidth\hsize\csname
@twocolumnfalse\endcsname
\begin{figure}
\noindent
\begin{minipage}{178mm}
\hspace*{20mm} \leavevmode \epsfxsize=138mm
\epsfbox{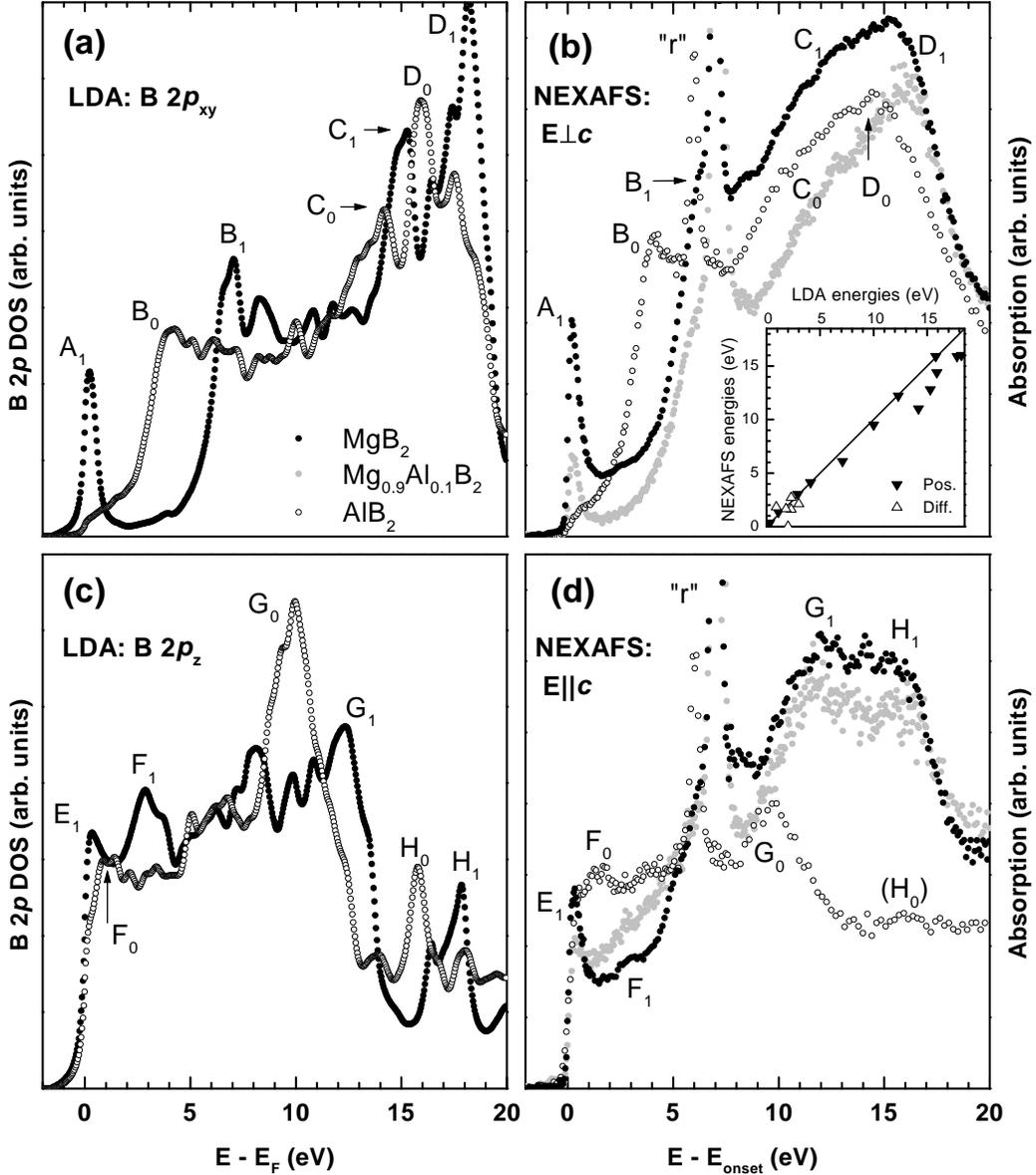} \caption{Orbital-specific electronic
structure of single-crystalline MgB$_2$ (black circles),
Mg$_{0.9}$Al$_{0.1}$B$_2$ (grey circles), and AlB$_2$ (open
circles): (a) Projected unoccupied density of states (DOS) with B
2$p_{xy}$ (``in-plane'') character from LDA band structure
calculations. (b) B 1$s$ near-edge x-ray absorption (fluorescence
yield detection) with the polarization of the incident radiation,
{\bf E}, oriented parallel to the base planes, probing the
unoccupied B 2$p_{xy}$ density of states. (c) As (a), but showing
the theoretical unoccupied DOS with B 2$p_z$ (``out-of-plane'')
character. (d) As (b), but with {\bf E}$\parallel$$c$, probing
the unoccupied B 2$p_z$ density of states. Inset: Comparison
NEXAFS $v$$s$.\ LDA\@. Full symbols denote peak positions; good
correspondence to the NEXAFS=LDA line indicates that correlation
effects are small. Open symbols denote the energy shifts between
AlB$_2$ and MgB$_2$ for all major peak pairs $X_0$/$X_1$
($X$=B$\dots$H); the clustering around 2 eV points to a
``semi-rigid'' band shift by that value.}
   \label{fig:NEXAFS}
\end{minipage}
\end{figure}]

\noindent detection with a typical probing depth of $\approx$500
{\AA}, using either a channeltron (SRRC) or Si(Li) detectors
(NSLS)\@. The necessary self-absorption correction was performed
\cite{SAC} using tabulated cross sections \cite{yeh}\@. While the
energy-dispersive Si(Li) detectors allowed direct suppression of
the mainly Mg/Al 2$p$-derived and energy-dependent background,
this background had to be numerically removed (using the same
tables) for the channeltron spectra. Some residual distortion
remains, and spectral weights for the latter, i.\ e., for MgB$_2$
and Mg$_{0.9}$Al$_{0.1}$B$_2$, cannot quantitatively be compared
to those for AlB$_2$\@. Spectra were taken at normal incidence
(polarization parallel to the crystal base plane, i.\ e., {\bf
E}$\perp$$c$) and at 60$^{\circ}$ grazing incidence; the {\bf
E}$\parallel$$c$ spectra were extrapolated. Energy calibration is
approximated by extrapolation from the O 1$s$ edge of a reference
system and yields a MgB$_2$ onset position near 186 eV\@.

LDA band structure calculations up to 20 eV above the Fermi level
$E_F$ were performed in the mixed-basis pseudopotential framework
\cite{meyer}, with 18/18/18 Monkhorst-Pack $k$ sampling in the
Brillouin zone (BZ) and using the tetraeder method for
determining the electronic density of states (DOS); further
details can be found in Ref.\ \cite{bohnen01}\@. The occupied
states (which are not observable in x-ray absorption) were
separated off, and convolution with a Gaussian of width 0.2 eV
roughly simulates experimental broadening effects, including
core-hole lifetime, multiphonon processes, and monochromator
resolution. In the absence of strong correlation and/or excitonic
effects the resulting B 2$p$ DOS should closely resemble the
experimental spectra.

Figure 1 presents the main results of this work. It is divided
into four panels (a)--(d): the left column [(a), (c)] contains the
theoretical results and the right column [(b), (d)] the
experimental ones; the upper row [(a), (b)] plots the in-plane
spectra, the lower row [(c), (d)] the out-of-plane spectra. Data
for $x$=1.0 (MgB$_2$) are depicted as black circles in all
panels, those for $x$=0.9 as grey circles, and those for $x$=0
(AlB$_2$) as open circles. To facilitate further discussion the
main features in the spectra are labeled A$\dots$H\@, with
subscripts ``1'' and ``0'' denoting features corresponding to
samples with or without Mg. A number of observations immediately
catches the eye and is discussed in the following:

(i) Most obvious is the strong asymmetry between in-plane and
out-of-plane spectra, illustrating the different character of B
$2p$ $\sigma$ and B $2p$ $\pi$ states, resp., and also
underlining the need for such a polarization- and thus
symmetry-resolved study.

(ii) In NEXAFS (but not in LDA), a sharp spike is observed around
6-7 eV above onset for both orientations (and for $x$=1.0 and 0.9
extending above the intensity region shown)\@. It is attributed
to one of the resonances identified in Ref.\ \cite{callcott011}
(and is thus labeled ``r'' in our Figure 1) \cite{spike}\@. This
feature does not affect the electronic structure near onset and
can thus be disregarded in the following.

(iii) The general agreement between theory and experiment is, on
all accounts, excellent: all major spectral features are
reproduced nicely, not only in their energy positions (Table I
provides a listing) but often also in their relative spectral
weight. For the {\em in-plane} spectra of MgB$_2$ the LDA results
(Fig.\ 1 (a)) show right above $E_F$ a prominent peak A$_1$
derived from a region of low dispersion along the $\Lambda$ line
in the BZ (the uppermost part of the almost fully occupied,
bonding $\sigma$ band derived from B $2p_{xy}$ states)\@. This is
followed by a region of very low (but non-zero) DOS until a steep
rise up to a shoulder B$_1$ occurs about 7 eV above $E_F$;
between 14 and 20 eV a large maximum D$_1$ is preceded by a
shoulder C$_1$ and followed by a steep decrease to low DOS\@.
NEXAFS data (Fig.\ 1 (b)) display the same feature A$_1$ right at
onset, very prominently and sharp. In fact, the width of this
feature, 0.7 eV, is exactly the theoretical value. Although the
spectral intensity right above A$_1$ is, in agreement with
theory, substantially reduced the suppression is not quite as
complete since the NEXAFS experiment does not exactly image the
DOS but starts developing an absorption jump. B$_1$ is partially
masked by the oxide-related peak ``ox'', but peak D$_1$ (with
shoulder C$_1$ before and steep decrease afterwards) reproduces
again nicely the averaged LDA structure in this energy range. The
in-plane spectra of AlB$_2$ exhibit for both LDA and NEXAFS a
very low DOS for about 2 eV above $E_F$ and onset, resp.\ --
there is no peak A$_0$ visible, suggesting that this feature
(which would correspond to A$_1$) is already occupied and
situated below $E_F$\@. Only after several eV the spectra rise to
a shoulder with small protrusion B$_0$ at about 4 eV above
$E_F$\@. D$_0$, again with preceding shoulder C$_0$ and
subsequent fall-off, is somewhat less pronounced compared to the
similar feature for MgB$_2$\@.

\noindent While NEXAFS for Mg$_{0.9}$Al$_{0.1}$B$_2$ in general
resembles that for MgB$_2$ quite closely it is conspicuous that
even at this small substitution level peak A$_1$ has already lost
much of its spectral weight, about 65\% compared to MgB$_2$\@.
Close inspection also indicates an A$_1$ width reduced by about
0.1 eV (less than all experimental broadening; the intrinsic width
reduction will thus be greater)\@. With this, and assuming that
the Al content injects 0.1 electrons per formula unit, one can
crudely estimate the B $2p_x$ DOS at E$_F$ to be $\lesssim$0.37
states/(eV$\cdot$atom); LDA estimates 0.10
states/(eV$\cdot$atom)\@. Obviously small additional substitution
will push A$_1$ below $E_F$ altogether, consistent with the quick
suppression of superconductivity reported for Al doping.

(iv) For the {\em out-of-plane} orientation LDA (Fig.\ 1 (c))
predicts, in contrast to the in-plane spectra, an almost constant
DOS for about 5 eV above $E_F$ for both MgB$_2$ and AlB$_2$\@.
Although the corresponding $\pi$ bonds give less rise to dramatic
DOS variations than the $\sigma$ bonds discussed in (iii) some
more features can be identified and exhibit, again, good
correspondence between theory and experiment. Differences worth
noting are very few: peak H$_0$ is almost undetectable in
experiment while H$_1$ is stronger in weight than in LDA; both
are, on the other hand, located 15 eV and more above $E_F$ where
precise correspondence between x-ray absorption and band
structure is less to be expected. Right at $E_F$, however, peak
E$_1$ for MgB$_2$ appears considerably stronger in NEXAFS than in
LDA; imperfections in mounting the small crystallites may have
introduced some angular uncertainty which could cause some
contribution from in-plane A$_1$ intensity.

\noindent The inset in Fig.\ 1 (b) further illustrates the good
agreement between LDA and NEXAFS: the peak positions (full
symbols) lie, up to 12 eV and more, very close to the straight
line denoting exact, 1:1 correspondence; the tendency visible at
higher energies towards reduced NEXAFS energies does not suggest
a significant amount of correlation. The open symbols depict the
corresponden-
\begin{figure} \noindent
\begin{minipage}{86mm}
\leavevmode \epsfxsize=86mm \epsfbox{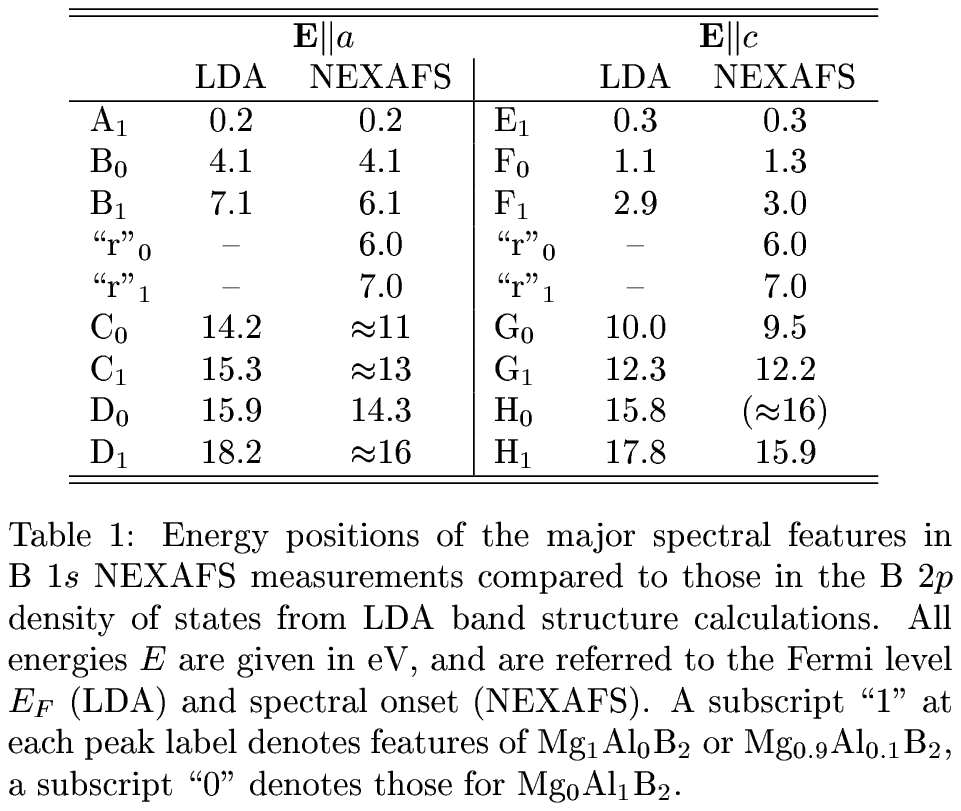}
\label{table1}
\end{minipage}
\end{figure}
\noindent ces between the peak pairs for AlB$_2$ and MgB$_2$,
like B$_0$/B$_1$, many of which are to some degree consistent
with a ``semi-rigid shift'' of the band structure \cite{an01} due
to the extra electron per formula unit introduced by Al replacing
Mg. The symbols cluster around a value of 2 eV for such a shift.

In conclusion, the orbital-resolved electronic structure of the
novel superconductor MgB$_2$ was studied with
polarization-dependent B 1$s$ NEXAFS on single crystals and was
compared to NEXAFS on single-crystalline
Mg$_{0.9}$Al$_{0.1}$B$_2$ and non-superconducting AlB$_2$ as well
as to the unoccupied B 2$p$ DOS obtained from LDA band structure
for both MgB$_2$ and AlB$_2$\@. The similarity between LDA and
NEXAFS is excellent for a large number of major spectral features
-- in particular, a B 2$p_{xy}$ $\sigma$ derived state at $E_F$
that is identified by theory to carry superconductivity shows up
very prominently in MgB$_2$ NEXAFS, is already considerably
reduced for Mg$_{0.9}$Al$_{0.1}$B$_2$, and is completely absent
(i.\ e., occupied) for AlB$_2$\@. Less drastic changes are
observed for the B $2p_z$ $\pi$ states\@. Systematics in peak
positions are consistent with a semi-rigid shift of the bands
expected from the extra electrons provided by Al. The close
correspondence between experiment and theory lends additional and
valuable support to further theoretical implications, including
the absence of strong correlation effects, the dominant effect of
holes in the B $2p$ $\sigma$ bands (as compared to the $\pi$
bands) for superconductivity, as well as its BCS-type mechanism
and electron-phonon origin, in this family of layered, covalent
compounds.

We thank C. T. Chen, S.-C. Chung, S. L. Hulbert, H.-J. Lin, G.
Nintzel, S. Tokumitsu, and H. Winter for generous support and
fruitful discussions. This work was supported by the German
Academic Exchange Service (DAAD) and the National Science
Council, Taiwan (NSC)\@. Research was carried out in part at the
NSLS, Brookhaven National Laboratory, which is supported by the U.
S. Department of Energy, Division of Material Sciences and
Division of Chemical Sciences, under contract number
DE-AC02-98CH10886.


\begin{thebibliography}{99}
\bibitem{B11}
Pure $^{11}$B was used to facilitate neutron scattering.
\bibitem{orient}
In-plane orientation is irrelevant due to the hexagonal symmetry.
\end{thebibliography}
\end{document}